\documentclass[atmp]{ipart_v1}

\usepackage{t1enc}
\usepackage[utf8]{inputenc}
\usepackage[english]{babel}

\usepackage{amsthm}
\usepackage{yfonts}

\usepackage{bbm}
\usepackage{bm}
\usepackage{mathrsfs}

\def \bea {\begin{eqnarray}}
\def \eea {\end{eqnarray}}
\def \nn {\nonumber}

\def \rr {\raise.35ex\hbox{\small $\prime$}\kern-.17em{\mbox{\large $\imath$}}}

\def \dels {\partial\kern-.6em /\kern.1em}
\def \As {{A\kern-.5em / \kern.5em}}
\def \Ds {D\kern-.7em / \kern.5em}

\def \ks {k\kern-.5em /}
\def \ls {l\kern-.5em /}

\newcommand{\be}[0]{\begin{equation}}
\newcommand{\ee}[0]{\end{equation}}

\newcommand{\ci}[1]{}

\newcommand{\ba}{\begin{eqnarray}}
\newcommand{\ea}{\end{eqnarray}}
\newcommand{\bal}{\begin{align}}
\newcommand{\eal}{\end{align}}
\newcommand{\bay}[1]{\left(\begin{array}{#1}}
\newcommand{\eay}{\end{array}\right)}

\newcommand{\hide}[1]{}

\numberwithin{equation}{section}

\theoremstyle{plain}

\begin{document}

\title[Discussion of Entanglement Entropy in Quantum Gravity]{Discussion of Entanglement Entropy in Quantum Gravity}

\author[Chen-Te Ma]{Chen-Te Ma}

\begin{abstract}
We study entanglement entropy in gravity theory with quantum effects. A simplest model is a two dimensional Einstein-Hilbert action . We use an $n$-sheet manifold to obtain an area term of entanglement entropy by summing over all background fields. A strongly coupled conformal field theory is expected to describe perturbative quantum gravity theory. Thus, we also use two dimensional conformal field theory to discuss a factorization of a Hilbert space. We find that a coefficient of a universal term of the entanglement entropy is independent of a choice of an entangling surface in two dimensional conformal field theory for one interval and also argue the result possibly be extended to multiple intervals. Finally, we discuss that translational invariance possibly be a necessary condition in a quantum gravity theory by ruing out a volume law of entanglement entropy.
\end{abstract}

\maketitle

\section{Introduction}
\label{1}
Quantum gravity is expected to obey a holographic principle, which is a description of a $D$ dimensional bulk gravity theory from a description of a $D-1$ dimensional boundary field theory. One candidate of the holographic principle is anti-de Sitter/Conformal field theory ($AdS/CFT$) correspondence. The evidences of the $AdS/CFT$ correspondence can be found from string theory, which is also a candidate of a perturbative quantum gravity theory. The $AdS/CFT$ correspondence offers a conjecture between a co-dimensional two minimum area term of a bulk $AdS$ metric and entanglement entropy of a boundary conformal field theory. 

A gauge invariant universal term of entanglement entropy in quantum field theory or a boundary conformal field theory is hard to define suffering from a decomposition of a Hilbert space \cite{Casini:2013rba,Ma:2015xes}. Because the problem of a factorization or decomposition of a Hilbert space appears near an entangling surface, it should be a ultraviolet problem \cite{Harlow:2015lma}. Hence, an ultraviolet complete quantum gravity theory possibly does not suffer from the problem. 

A geometric co-dimensional two minimum area from a bulk metric enhances an area law of entanglement entropy in quantum gravity.
A first generic mathematical proof for the area law of entropy is \cite{Casini:2003ix} based on the translational invariance, Poincaré symmetry, causality and finite entropy. The result is a sum of a constant term and an area term \cite{Casini:2003ix}. Here, we want to discuss the area law  of entanglement entropy by using translational invariance without too many restricted conditions.

Our goal of the paper is to discuss the entanglement entropy in a gravity theory with quantum effects. To provide an explicit example of an area term in a gravity theory, we consider a two dimensional Einstein-Hilbert theory and compute entanglement entropy. For studying a factorization of a Hilbert space in a perturbative quantum gravity theory, we discuss two dimensional conformal field theory by using a modular symmetry \cite{Casini:2004bw} and uniqueness of the mutual information \cite{Huang:2016bkp} to prove uniqueness of the coefficient of a universal term of the entanglement entropy or center charge \cite{Ohmori:2014eia}, which shows that a factorization of a Hilbert space is not problematic in two dimensional conformal field theory. Entanglement entropy of $N$ intervals is also reproduced from a geometric method or a holographic method in two dimensional conformal field theory \cite{Faulkner:2013yia}. Any bulk geometry should not be affected by a choice of the entangling surface so this should also indicate that a coefficient of a universal term of entanglement entropy should be independent of a choice of the entangling surface for generic multiple intervals. Finally, we argue that translational invariance is a necessary condition to rule out a volume law of entanglement entropy in a perturbative quantum gravity theory. 

\section{Entanglement Entropy in a Two Dimensional Einstein Gravity Theory}
\label{2}
An action of a two dimensional Einstein-Hilbert theory is defined by
\bea
S_{EH}=-\frac{1}{16\pi G}\int d^2x\ \sqrt{\det{g_{\mu\nu}}}R=-\frac{1}{4G}\chi,
\eea
where $\chi\equiv 2-2g$, $g$ is the genus, $G$ is the Newton constant, and
\bea
R_{\mu\nu}&\equiv&\partial_{\delta}\Gamma^{\delta}_{\nu\mu}-\partial_{\nu}\Gamma^{\delta}_{\delta\mu}
+\Gamma^{\delta}_{\delta\lambda}\Gamma^{\lambda}_{\nu\mu}
-\Gamma^{\delta}_{\nu\lambda}\Gamma^{\lambda}_{\delta\mu}, 
\nn\\
 \Gamma^{\mu}_{\nu\delta}&\equiv&\frac{1}{2}g^{\mu\lambda}\bigg(\partial_{\delta}g_{\lambda\nu}+\partial_{\nu}g_{\lambda\delta}
-\partial_{\lambda}g_{\nu\delta}\bigg),
\nn\\
\eea
\bea
 R\equiv g^{\mu\nu}R_{\mu\nu}.
\eea 
We also label the Greek letters as the spacetime indices. The entanglement entropy $S_{EE}$ is computed by an $n$-sheet method \cite{Gromov:2014kia} and the result is
\bea
S_{EE}=-\sum_ip_i\ln p_i+\frac{\langle N\rangle}{2G}+\cdots,
\eea
where
\bea
p_i\equiv\frac{e^{\frac{1}{4G}\chi_i}}{\sum_{\chi}e^{\frac{1}{4G}\chi}},
\eea
$2N$ is a number of ramification points and $\langle N\rangle$ is the expectation value of $N$. We used the Riemann-Hurwitz theorem
\bea
\chi_n=n\chi-2N(n-1),
\eea 
in computation of entanglement entropy, where $\chi_n$ is the Euler characteristic in the $n$-sheet manifold. The path integral is defined by summing over all manifolds with different genus and numbers of ramification points. When we use the $n$-sheet method, the degeneracy of closed manifolds possibly be modified by $n$. Here, we do not consider the changing of the degeneracy of closed manifolds so we may have $\cdots$ in entanglement entropy. This expression is still interesting because the result already contains the sum of classical Shannon entropy and the area term, which also be seem as quantum extension of two dimensional finite entropy \cite{Casini:2003ix}. Because two dimensional Einstein-Hilbert theory also has conformal symmetry \cite{Casini:2011kv}, we could obtain consistent results for $N$ intervals, $\alpha+\beta N$, where $\alpha$ and $\beta$ are constants, as in two dimensional finite entropy \cite{Casini:2003ix}. We define that an analogous area term in two dimensional theories is a number of ramification points. For each interval, we have two ramification points. Thus, this example shows an expected area term for an ultraviolet complete quantum gravity theory. 

\section{Two Dimensional Conformal Field Theory}
\label{3}
The translational invariance and strong subadditivity \cite{Ma:2015xes,Huang:2016bkp,Araki:1970ba,Lieb:1973cp,Casini:2014aia,VanAcoleyen:2015ccp} can give
\bea
S_A(l_A)+S_B(l_B)\ge S_{A\cup B}(l_{A\cup B})+S_{A\cap B}(l_{A\cap B}),
\eea
in which $S_A$ is the entanglement entropy of a region $A$, $S_B$ is the entanglement entropy of a region $B$, $S_{A\cup B}$ is the entanglement entropy of a region $A\cup B$, $S_{A\cap B}$ is the entanglement entropy of a region $A\cap B$, $l_A$ is the length of the region $A$, $l_B$ is the length of the region $B$, $l_{A\cup B}$ is the length of the region $A\cup B$ and $l_{A\cap B}$ is the length of the region $A_{\cap B}$. 

We could consider the Poincaré symmetry or boost symmetry to constraint the length of the systems for one interval in two dimensional quantum field theories \cite{Casini:2004bw} to obtain
\bea
l_{A\cup B}\cdot l_{A\cap B}=l_{A}l_B.
\eea
Then we use a modular transformation
\bea
x\rightarrow\frac{ax+b}{cx+d}, \qquad ad-bc=1,
\eea
in which parameters $a$, $b$, $c$ and $d$ are constants, to know
\bea
&&F\bigg(\frac{(u_2-u_3)(u_1-u_4)}{(u_1-u_3)(u_2-u_4)}, \frac{(v_2-v_3)(v_1-v_4)}{(v_1-v_3)(v_2-v_4)}\bigg)
\nn\\
&=&S_A\bigg(\sqrt{(u_1-u_3)(v_1-v_3)}\bigg)+S_B\bigg(\sqrt{(u_2-u_4)(v_2-v_4)}\bigg)
\nn\\
&-&S_{A\cap B}\bigg(\sqrt{(u_2-u_3)(v_2-v_3)}\bigg)-S_{A\cup B}\bigg(\sqrt{(u_1-u_4)(v_1-v_4)}\bigg),
\nn\\
\eea
where $u_i=t_i+x_i$ and $v_i=t_i-x_i$ are null coordinates. The form of the entanglement entropy could be uniquely determined as:
\bea
S_A=k_1\ln l_A+k_2,
\eea
where $k_1$ and $k_2$ are constants.
Thus, the form of the entanglement entropy is uniquely determined from translational invariance, strong subadditivity, a boost symmetry and a modular symmetry \cite{Casini:2004bw}. The algebraic method could include different choices of an entangling surface of entanglement entropy. Now it is not enough to show uniqueness of universal terms of entanglement entropy because different choices of an entangling surface may give different coefficients ($k_1$, $k_2$). To show the result, we use the method \cite{Huang:2016bkp} to show that mutual information
\bea
k_1\ln\frac{l_Al_B}{l_{A\cup B}}+k_2
\eea
is unique if each region ($A$, $B$ and $A\cap B$) is one interval and we only consider different choices of an entangling surface on the boundary of $A\cup B$. Then we could know that the coefficient of the universal term of entanglement entropy for one interval is unique. For some examples of multiple intervals, the universal terms of the entanglement entropy are also unique, which could also be deduced from \cite{Huang:2016bkp}.

To extend our discussion to generic multiple intervals, we could use a higher genus Riemann surface, which is obtained from quotient of a complex plane by a discrete subgroup of $SL(2, C)$, to determine a universal term of entanglement entropy of $N$ intervals \cite{Faulkner:2013yia}. The result of entanglement entropy can give a consistent result as in two dimensional finite entropy \cite{Casini:2003ix}. When a regularization parameter is small and finite, a dominant term of entanglement entropy dependence on a regularization parameter. Hence, the dominant term is consistent with a result of two dimensional finite entropy. Thus, we conclude that a coefficient of a universal term of entanglement entropy of $N$ intervals in two dimensional conformal field theory should be unique because a bulk geometry should not be modified by a choice of an entangling surface. 

We know that entanglement entropy of a boundary field theory is not universal for $N$ intervals in two dimensional conformal field theory so it depends on different regularization methods \cite{Calabrese:2010he} for a trivial choice (without removing any operators from an entangling surface) \cite{Casini:2013rba}. Here, we show that a coefficient of a universal term of entanglement entropy are not modified when we choose different entangling surfaces.

\section{Non-Volume Law of Entanglement Entropy}
\label{4}
We want to rule out a volume law of entanglement entropy to find physical principles in an ultraviolet complete perturbative quantum gravity theory. If a theory has translational invariance and satisfies subadditivity \cite{Ma:2015xes}, the density of entanglement entropy must vanish when we take an infinite volume limit \cite{Casini:2003ix}. To argue that a volume law of entanglement entropy does not exist in a quantum system, we also consider zero temperature and an infinite spatial size without mass scales, except for a cut-off, then a volume term of the entanglement entropy in a region with a volume $V$ must be proportional to $V/\epsilon^{D-1}$, where $\epsilon$ is the regularization parameter and $D$ is dimensions of spacetime. Thus, it is easy to find that a density of entanglement entropy should be divergent when the regularization parameter is small and the volume $V$ is infinite because we expect that finite quantities or the density of entanglement entropy should not be modified by taking $\epsilon\rightarrow 0$. Thus, we argue that a non-volume law of entanglement entropy at zero temperature in a quantum system with an infinite spatial size and without mass scales, except for the cut-off, needs translational invariance. If we consider an ultraviolet complete perturbative quantum gravity theory or a conformal field theory, time translational invariance already implies unitary of a quantum system to guarantee the strong subadditivity. An ultraviolet complete perturbative quantum gravity is also expected to be scale invariant so our discussion without mass scales, except for the cut-off, does not lose generality when we discuss an ultraviolet complete perturbative quantum gravity theory. We first assume that the area law seems to be a necessary condition in a quantum gravity theory so we use translational invariance to rule out a volume law of entanglement entropy.

\section{Conclusion}
\label{5}
An area law of entanglement entropy could be obtained from the $AdS/CFT$ correspondence \cite{Naseh:2016maw,Allahbakhshi:2013rda}. We demonstrated the area law of entanglement entropy in a two dimensional Einstein-Hilbert action by summing over all background fields. The problems of a factorization of a Hilbert space possibly disappear in an ultraviolet complete perturbative quantum gravity theory, which is demonstrated by two dimensional conformal field theory. Finally, we show that the volume terms of the entanglement entropy cannot not exist when a quantum system with infinite size has translational invariance at zero temperature and the quantum system does not have any mass scale, except for a cut-off. Thus, our studies should enhance existence of the area law of entanglement entropy in a quantum gravity theory. In other words, constructing an ultraviolet quantum gravity theory should not violate an area law of entanglement entropy.

\subsection*{Acknowledgments}

The author would like to thank Horacio Casini, Daniel Harlow, Song He, Kazuo Hosomichi, Xing Huang, Yu-Tin Huang and Ling-Yan Hung for their useful discussion. Especially, the author would like to thank Nan-Peng Ma for his suggestion and encouragement.




\providecommand{\href}[2]{#2}

\address{
Department of Physics, Center for Theoretical Sciences, \\
Center for Advanced Study in Theoretical Sciences,\\
 National Taiwan University, Taipei 10617, Taiwan,
R.O.C..\\
\email{yefgst@gmail.com}\\
}

\end{document}